**PAPER • OPEN ACCESS**

# Internal process: what is abstraction and distortion process?

View the article online for updates and enhancements.

**Related content**

- Biophysics of the Senses: Electrical properties of the body
  T D Presley

- Spiral Structure in Galaxies: The classification of galaxies
  M S Seigar

- Introducing geometry concept based on history of Islamic geometry
  S Maarif, Wahyudin, A Raditya et al.

# Internal process: what is abstraction and distortion process?

**F R Fiantika[1,*], I K Budayasa[2] and A Lukito[2]**

[1,2]Department of Mathematic Education, Universitas Negeri Surabaya, Indonesia
[1]Department of Mathematic Education, Universitas Nusantara PGRI Kediri, Indonesia

*Corresponding author: fentfeny@gmail.com

**Abstract.** Geometry is one of the branch of mathematics that plays a major role in the development of science and technology. Thus, knowing the geometry concept is needed for students from their early basic level of thinking. A preliminary study showed that the elementary students have difficulty in perceiving parallelogram shape in a 2-dimention of a cube drawing as a square shape. This difficulty makes the students can not solve geometrical problems correctly. This problem is related to the internal thinking process in geometry. We conducted the exploration of students' internal thinking processes in geometry particularly in distinguishing the square and parallelogram shape. How the students process their internal thinking through distortion and abstraction is the main aim of this study. Analysis of the geometrical test and deep interview are used in this study to obtain the data. The result of this study is there are two types of distortion and abstraction respectively in which the student used in their internal thinking processes.

## 1. Introduction

Geometry is a branch of mathematics concerned with shape, size, a relative position of figures, and the properties of space. Geometry is an important area of mathematics which can be found in the natural world as well as in virtual areas of human creativity and ingenuity. Geometric representations can help students make sense of area and fraction, histograms, and scatterplots, can give insight about data, and coordinate graph, can serve to connect geometry and algebra [1]. It is important across several disciplines including engineering and the basic sciences enables an individual to visualize, edit, reorganize and generalize facts and is required in diverse workplace settings, such as mechanical engineering, pilot training, and scientific crime scene investigation [2]. Although the geometrical object is really close to human activity, students have difficulty in imagining and conceptualizing a 3-dimensional (3D) geometrical object particularly from its 2-dimensional (2D) representation in the plane [3]. There is a need to explicitly interpret and utilize drawing 3D objects conventions, otherwise, students may misread a drawing and do not understand whether it represents a 2D or 3D object.

　　Representation have a play role in this season. There are include abstraction and distortion. Abstraction and distortion are like reality in one peace our lived, when we "see" only some of what lies in front of us,"hear" only part of the noises in our vicinity and ignore part of information. The authors found that as many as 60% students from three different school in a city still have difficulty in classifying four sides object such as square, rectangular, parallelogram and rhombus. Four fifth grade of elementary students were asked to draw a net of a given cube. Although the students can do the task correctly the model of cube nets drawn by the students is the same.

　　The results of interviews from the students said that students can draw the cube nets according to what the teacher taught. The students have difficulty in constructing different cube models and saying that there is no other model of cube nets to the same cube. Hence, the representation of students in the process of exploring the subject mentioned is required.

## 2. Theoretical Framework

Representation is process of thinking, there is certainly a useful concept in the basis for explanation something (internal represent: desire, belief, etc; external; image, graph, etc). Representations are divided into two categories: internal representations and external representations [4, 5, 6, 7, 8]. External representations refer to physical constructs (e.g., algebraic expressions, graphs, or diagrams written or sketched on a paper) that the teachers use to illustrate mathematical ideas to their students [5]. They are spoken, written or some sort of visible entities. Internal representations are mental constructs that the individuals develop through their interactions with and upon the external representations [6, 8].

The representation is internal process, therefore it is needed visualization to make the ideas to be concrete and someone understand to our ideas. There are included abstraction and distortion in the representation process. Abstraction is sensory characteristics of embodied objects are ignored and only the intelligible ones are kept, this is specific mental process in which new idea by conceptions are formed by considering several objects or ideas and omitting the features that distinguish them [9]. Abstraction is a process that reduces the information or observation of phenomenon, and retain information which is relevant with a purpose of individual.

It is individual mental process which is compared two objects characteristics or more and collected the relevant information. We can said that an abstraction is remove the real object form, there is omitting the conditions because an abstraction involves the omission of a truth. In the abstraction process, we do not describe the system in it is entirety, so there is not telling the whole truth. An abstraction is include "freedom" a distorted, irrelevant information is not to use. Omission necessarily results in distortion [10]. Thus, while abstraction can result in the distortion, abstraction and distortion are very different in each case. Indicator of internal representations can be shown in Table 1.

Table 1. Indicator of internal representations

| Component | Indicator |
|---|---|
| Abstraction | Imagine the shape or position of an object geometry that is viewed from a certain angle and dimension. |
| Distortion | 2. Imagine relationships between objects |
|  | 3. Imagine the direction of the movement of an object |
|  | 4. Imagine how the direction of the movement of an object |

### 3. Methods
Explorative descriptive through qualitative approach was used in this study.

*3.1 Subject*
The subjects of this study were the 5th grade of visual learning style of both boy and girl students because geometry is closely related to visualization. The 5th-grade students were chosen because they have learned space matter in school [11] and still in concrete operation phase [12], also it is important to study this subject in the earliest of cognitive development. The student selected carefully based on the steps: 1) The students were divided into three learning style ( visual, auditory and kinesthetic) based on the learning style test [13], 2) The visual learning style student were given mathematical test to classify into low, middle and high ability in mathematics, 3) a boy and a girl were chosen from middle ability group. Numerous studies show that the gender influence students' performance and concept in mathematics [14, 15] and that boys do better in math [14] particularly in space [7, 14] than the girls. Besides exploring the spatial thinking of students, the author also distinguishes between boy's and girl's work to provide comprehensive information.

*3.2 Technique*
The subjects work in solving geometry test were explored and analyzed through an unstructured interview where the interviewer follows the interviewees narration and generates questions spontaneously based on his or her reflections on that narration. A deep interview conducted to the subject regarding what they thought, done, written and spoken while doing the test. The audio and video format were used to record subjects along the research process from solving geometry test until interview section. All data are selected according to the need to answer research questions.

*3.3 Instrument*

There were two instruments used in this studyto explore the student's internal process in geometry concept with using cube object. The first instrument contained statements about concepts which the students could explore the cube nets from the given cube object and reasons to strengthen their justification. The second were student learning style test to classify student learning style and mathematics ability test to classify students into a low, middle and high level that shown in Table 2.

**Table 2.** Classification of student's learning style

| Gender | Learning Style | | | | | | | | |
| --- | --- | --- | --- | --- | --- | --- | --- | --- | --- |
| | Auditory | | | Visual | | | Kinesthetic | | |
| | Low | Middle | High | Low | Middle | High | Low | Middle | High |
| Boy | 0 | 4 | 0 | 1 | 2 | 1 | 2 | 2 | 2 |
| Girl | 2 | 4 | 2 | 0 | 3 | 1 | 1 | 3 | 0 |
| Total | 30 | | | | | | | | |

The authors use learning style instruments developed by Gunawan [13] to classify student who has visual learning style. There are 36 items statement given to the students for each 12 statement items for visual, auditory and kinesthetic learning styles, respectively. The mathematical ability test used in this study was a mathematical problem adopted from the national examination. The 10 questions were chosen according to the 5th-grade class material.

Four validity test data were used in this study: 1) credibility test (internal validity), 2) test of dependency (reliability), 3) confirmation test (objectivity) and 4) transferability test. The credibility test of the data used in this study is triangulation time and triangulation method. To sum up, intensive field study also conducted in this research i.e. discussing with colleagues, checking the reference sufficiency, reading the literature and performing member checks.

## 4. Result and discussion

*4.1 The similarity and differenceinternal representations betweenboy and girl*

Internal activity begins when the senses of the subject capture the cube model that is in front of him, the subject was looking at the model of the cube in front of him. The subject encoded the information obtained when viewing the cube model into his mind, the subject attributing information possessed to the current information that is the subject of reconstructing the cube into cubes of the cube in his mind.

Representation internal shown 1) when the subjects silence before deciding on something in speech or act. There have not happened abstraction and distortion 2) The subjects both boy and girl imagine opening the cube model until a cube net is obtained. This has meaning that they used abstractions, they omit differences characteristics between real and abstract cubes and used the relevant characteristics of real and abstract cubes. Real and abstract cubes develop by similar entities, they have six squares, they have similar orientation as frame reference and they are in 3-dimensional. Distortion is abstraction result. They ignored differences abstract and real object, they ignored how to construct abstract and real object. 3) imagining transforming dimensions of transforming 3-dimensional objects into 3-dimensional abstract object. This has meaning that they used abstractions, they omit differences characteristics between real and abstract cubes and used the relevant characteristics of real and abstract cubes. Real and abstract cubes develop by similar entities, they have six squares, they have similar orientation as frame reference and they are in 3 Dimension. Distortion is abstraction result. They ignored differences abstract and real object, they ignored how to construct abstract and real object . 4) Imagine performing 3-dimensional abstract object to a 2-dimensional abstract. This has meaning that they used abstractions, they omit differences characteristics between cube net's and abstract cubes and used the relevant characteristics. The abstract cube net's and abstract cube developed by six squares, sides of abstract cube have similar disposition with the abstract cube net's. They used abstract cube as reference object to develop abstract cube net's. Distortion in this process are they ignored differences form of cube and cube net's, they ignored differences of dimensional, ignored how to construct cube and cube net's. 5) Imagine performing a 2-dimensional abstractto a 2-dimensional image object. This has meaning that they used abstractions, they omit differences characteristics between real and abstract cube nets and used the relevant characteristics of real and abstract cube nets. Real and abstract cube nets develop by similar entities, they have six squares, they have similar orientation as frame reference and they are in 2 Dimension. Distortion is abstraction result. They ignored differences abstract and real object, they ignored how to construct abstract and real object. 6) Imagine performing mental manipulation of imagining matching sides to a square cube model with squares on cube nets, using an analogy for constructing sketches of cube nets. This shown abstraction process. Distortion is ignore physical form of objects. Table 3 shows the explanation of students related to how they imagine the cube object to cube nets.

Table 3. The differences internal representation process boy's and girl's

| Internal Representation process | Boy | Girl |
|---|---|---|
| Abstraction | Imagine object by analogy | Imagine phenomenon object |
| | Compared physical form of object | Compared physical form of object and other characteristics |
| | Imagine flips object | Imagine flips object and rotate object 90 degrees |
| Distortion | Using abstract object | Using abstract object and sketch |
| | Ignored role play of moving object | Ignored direction of moving |
| | Ignored numerical of performing process | Ignored phenomenon characteristics which they have choose |
| | Ignored detail transforming process | |
| | Ignored differences physical objects which they are used | |

Question : Can you tell me, how did you draw a sketch of cube nets based on the given dotted cube model ?
Boy : Firstly, I imagined opening the model of this cube, like a car changes into a transformer robot until the shape of the image of the nets made, then I make the net's sketch...
Girl : ... I try to start imagining the letters or anything I've ever seen which can be made with 6 pieces of a square. thus, if it folded then can construct a cube, ...
In another interview
Boy : First I imagine turning the cube model into a net like something I've ever seen, interesting, and it looks like it can be made sketches,…

## 5. Conclusion
In this research we have obtained research results that reveal the internal representations of elementary students in solving geometric problems based on gender. Generally, they have differences thinking steps; boy's has thinking steps; try to make cube nets and start to make it similar with their experience before, they match it to his phenomenon, and sketch it. Girl's has thinking steps; the beginning steps are finding the phenomenon which it is possible to make to cube nets, and sketch it. Researchers found 1) analogy as process solving of problem 2) there are include two types of abstraction; process and product. Abstraction as process is mental process, it is happen in human mind. There have collected the similar characteristics and ignore irrelevant characteristics. An object is abstract, it is the referent of an abstract idea, i.e., an idea formed by abstraction. It is mean abstraction as product. 3) There two types of distortions; process and product. Distortion as process is ignored how to get the objects and distortion as product is used to build new idea. 4) Abstraction and distortion is duo, there do an abstraction in other side distortion is follow it.

**Acknowledgement**
The author thanks to Universitas Nusantara PGRI Kediri and Universitas Negeri Surabaya for supporting this research. Also thanks to Indonesia endowment fund for education for supporting the grant.